\documentclass[a4paper]{mem}
\usepackage{natbib}
\usepackage{graphicx}
\begin{document}
   \title{A Probe of the Matter  Content of Quasar Jets
%\thanks{this is a place for a title footnote}
}

   \author{M. Georganopoulos \inst{1,2}, 
   D. Kazanas \inst{2},\\ 
           E. S. Perlman \inst{2}
          \and
          F. W. Stecker \inst{1} %\fnmsep
%\thanks{this is a place for placing a footnote in the author field }
}

   \institute{Department of Physics, Joint Center for Astrophysics, University of Maryland-Baltimore County, 1000 Hilltop Circle, Baltimore, MD 21250, USA.   \email{markos@milkyway.gsfc.nasa.gov}\\ 
              \and Laboratory for High Energy Astrophysics, 
NASA Goddard Space Flight Center, 
Code 661, Greenbelt, MD 20771, USA 
             }

   \abstract{  
We propose a method for estimating the matter content of quasar jets 
which exhibit {\sl Chandra -}  detected knots in their kpc scale jets. 
The method relies on measuring the component of the Cosmic Microwave  
Background (CMB) radiation that is bulk-Comptonized (BC) by the cold electrons 
in the relativistically flowing jet.
 We  apply our method to 
PKS 0637 -- 752, a superluminal quasar with an one -- sided 
 {\sl Chandra} -- detected large scale jet. 
What makes this source particularly suited for such a procedure is  the 
absence of  significant  non-thermal jet emission in the {\sl bridge},
 the region between 
the core and the first bright  knot, guaranteeing that most of the electrons
in the bridge are cold, and  
leaving the BC scattered CMB radiation as the 
only  significant source of photons in this region.
At $\lambda=3.6-8.0 \; \mu m$, 
the most likely band for the BC scattered emission to appear,
 the   angular resolution of {\sl Spitzer} ($\sim 1''-3''$) is considerably 
smaller than 
the bridge of PKS 0637 -- 752   ($\sim 8''$), making it possible to both 
measure and resolve this emission.
   }
   \authorrunning{M Georganopoulos et al.}
   \titlerunning{The matter content of quasar jets}
   \maketitle
%
%________________________________________________________________

\section{Introduction}

The composition of extragalactic jets continues to remain elusive.
A number of attempts (e.g. Reynolds et al. 1996, Wardle et al. 1998) have 
been made over the years toward measuring, or at
the least constraining, the matter content of jets, and in particular the 
fraction of kinetic energy stored in protons and low energy or cold leptons, 
whose low radiative efficiencies fail to provide direct evidence of their
presence.

A direct estimate of the cold lepton content of  blazar jets  was proposed  by
Sikora \& Madejski (2000):
The observed  non-thermal blazar emission is thought to be produced at 
distances $\sim 10^{17}-10^{18}$ cm  from the central engine; 
the jet leptons providing the blazar emission at 
these distances need to be transported 
practically cold by a relativistic flow of bulk Lorentz factor $\Gamma \sim 10$
from the black hole vicinity to the blazar emission site; as these cold jet 
leptons  propagate through the blazar broad line region (BLR) they would  
Compton -- scatter the BLR optical-UV photons  to energies 
$ \sim 1$ keV, to produce a black -- body type hump in their X-ray spectra. 
The fact  that such a feature is not observed in the inverse-Compton
 dominated X-ray spectrum of blazars, led the above authors to conclude
that the jet power  is carried mainly by protons,
 although cold leptons  dominate the number of particles in the jet. 

While this idea is well founded and appealing, concrete answers
are hindered by unknowns such as the  distance  at which the jet is formed, 
its sub-pc scale opening angle and the actual photon energy density of 
the BLR, as well as by the presence of a strong X-ray non-thermal continuum 
that apparently could  ``hide" the proposed bulk-Comptonized component.

%                                     Two column figure (place early!)
%______________________________________________ Gamma_1 (lg rho, lg e)
   \begin{figure*}
   \centering
   \resizebox{3.in}{!}{\rotatebox[]{0}{\includegraphics{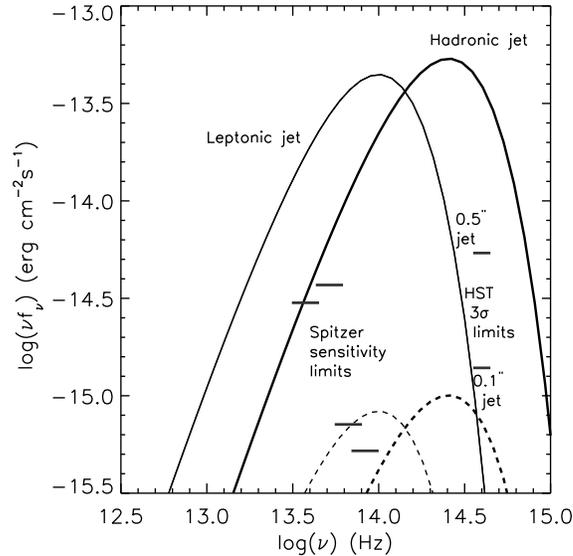}}}
   \caption{The BC emission for an $e^\pm$ ($e-p$) jet composition of
 PKS 0637-752  is plotted with a solid thin (thick) line for case A,  
in which the lepton power $L_{lept}$ 
required in the knot is provided by the cold leptons in the beam. 
The dashed lines correspond to case B, in  which the jet provides
 simply the number of  leptons needed in the knot, with the thin (thick) line 
representing an $e^\pm$ ($e-p$) jet composition. The {\sl Spitzer} 
sensitivity limits and  existing $3\sigma$ {\sl HST} limits from Schwartz et al. (2000) assuming a $0.1''$ or a $0.5''$
 jet radius are also shown. }
              \label{eikona1}%
    \end{figure*}

%\section {The method}
\section{Using the CMB as a seed photon source}

Arguments  based on the BC emission 
can be applied  to any 
astrophysical site involving relativistic flows. One can then obtain more 
concrete conclusions provided that the flow geometry
 and the target photon density
are better determined. Such a site is presented by the
 large scale  jets of {\sl Chandra} -- detected
superluminal quasars such as PKS 0637--752 (Schwartz et al. 2000;
 Chartas et al. 2001).
The jet of  PKS 0637--752, which  is resolved and is  found to be  well 
collimated, propagates through a well understood photon field:
the CMB. The source  exhibits radio, optical, and X-ray  emission from the 
quasar core and then from well separated knots along the jet at angular 
distances $\sim 8'' $.
The fact that the bridge, the region between the core and the first  
knot WK7.8,  radiates weakly in radio, optical, 
and X-ray energies  is very important because:  
({\it i}) it shows that most of the leptons 
propagating  in the bridge are cold  
 and ({\it ii}) it provides a region free from
 unwanted contamination by unrelated  broad 
band non-thermal radiation. 

The luminosity $L_{BC}$ of the BC emission depends on the power 
$L_e$ carried by cold
leptons in the bridge, the length $l$ of the bridge, 
the bulk Lorentz factor $\Gamma$ of the flow, 
and the angle $\theta$ formed between the jet axis and the line of sight.
It can be  shown (Georganopoulos et al. 2004) that
\begin{equation}
L_{BC}\approx 1.4 \;10^{-4} \, l_{100 Kpc}  \Gamma_{10}^3 (1+z)^4  L_e,
\end{equation}
where $z$ is the redshift of the source and we have assumed  the typical for superluminal sources  $\theta=1/\Gamma$.
The BC component  peaks in the IR regime  
\begin{equation}
\nu_{BC}\approx 4 \; 10^{13} \; \Gamma^2_{10} \; {\rm Hz}, 
\end{equation}
regardless of $z$. The BC emission, therefore, requires an estimate
of the jet power and kinematics. These can be provided
by the spectrum and luminosity of the {\sl Chandra} --
detected knots, once the knot X-ray emission mechanism has been established. 

\subsection{The X-ray emission mechanism}

Schwartz et al. (2000) noted that the X-ray emission  from knots 
at a projected distance of $\sim 100$ kpc from the core of  PKS 
0637 -- 752 is
part of a spectral component separate from the  synchrotron  radio-optical
emission and it  is too bright to be explained through  synchrotron self 
Compton SSC emission from electrons in energy  equipartition with the jet 
magnetic field. 
Tavecchio et al. (2000) and Celotti et al. (2001) argued that 
the X-ray emission is due to external Compton (EC) scattering of CMB 
photons off 
relativistic electrons in the jet, provided that the jet flow is sufficiently 
relativistic ($\Gamma \sim 10$) to  boost  the 
CMB energy density in the flow frame (by $\Gamma^2$) to the level 
needed  to reproduce the observed X-ray flux. 
This was the first suggestion
that powerful jets retain significantly
relativistic velocities at large distances from the core, a very important
feature because it boosts the level of the  anticipated 
BC emission by $\sim \Gamma^2$. We adopt here this 
interpretation of the X-ray emission; the reader can find a discussion of the 
alternatives  in Georganopoulos et al. (2004).

\subsection{Minimum power conditions}

A  set of  constraints for the jet power and beaming based 
on multiwavelength observations of knots has been 
presented by Dermer and Atoyan (2004, hereafter DA).
These authors model the knots as homogeneous sources moving with a 
Lorentz factor 
$\Gamma$ at an angle $\theta$ to the line of sight.
The knot matter content  is described through 
 the ratio of  power carried by  
protons to  power carried by a power law lepton distribution.
Assuming that the X-rays are due to EC scattering off the CMB and that 
$\delta=\Gamma$, DA  calculate  
(their Eq. (12))  the Doppler factor $\delta_{min}$  
that minimizes the power  needed in the knot to produce the X-ray flux.
Because the 
minimized quantity is the total knot power and {\sl not} the power in relativistic
electrons and magnetic field only, {\sl $\delta_{min}$ depends 
 on the matter content of the jet, and it is higher for $e-p$ jets}
(Georganopoulos et al. 2004).

\subsection{The BC emission of PKS 0637-752}

Using  the formalism of DA we derive    minimum
power Doppler factors $\delta_{min}=17.4$ for a $e^\pm$ composition 
and   $\delta_{min}=27.8$ for an $e-p$ composition of knot WK7.8.
These    correspond to a jet 
minimum power $L_{min}=9.7 \times  10^{45} $ erg s$^{-1}$ for a $e^\pm$
jet and  $L_{min}=6.3 \times  10^{46} $ erg s$^{-1}$ for an $e-p$ jet.
The  corresponding lepton power is  $L_{lept}=3.7 \times 10^{45} $ erg s$^{-1}$
 for a $e^\pm$
jet and  $L_{lept}=6.8 \times  10^{44} $ erg s$^{-1}$ for an $e-p$ jet;
these numbers are only weakly affected by our choice of the low energy cutoff
$\gamma_{min}$ of the electron power law
as long as $\gamma_{min} >  10$, a condition imposed by the 
requirement that  EC does not overproduce the observed  optical flux.
Assuming $\theta=1/\Gamma$, at $z=0.651$
 $1''$ corresponds to $6.9$ Kpc
and  the actual bridge length
 is $l \approx $ 930 Kpc for a $e^\pm$ jet and $l \approx  1.5$
 Mpc for an $e-p$ jet. 
We calculate the BC flux
 for an $e-p$ and an $e^{\pm}$ composition
for two cases.

{\sl Case A:} 
The lepton
power $L_{lept}$ required in the knot is provided by the cold leptons in the 
beam ($L_{e}=L_{lept}$). This requires that only a minority of the leptons
get accelerated in the knot, and it is an optimistic estimate of the 
anticipated BC emission.

{\sl Case B:} The most conservative case for the anticipated
 BC emission, according to which the jet provides simply the number of 
leptons needed in the knot, and the leptons are accelerated in the knot using
energy exclusively from other agents such as  the magnetic field and/or 
the jet hadrons.

As can be seen in Figure \ref{eikona1}, in case A 
the emission for  a $e^\pm$ jet peaks at mid IR energies,
while that for an  $e-p$ jet peaks at near IR - optical energies. 
For both compositions  the anticipated mid IR flux is above the {\sl Spitzer} 
sensitivity limits; the $e-p$ case however violates the {\it HST} 3$\sigma$
detection limits  for both a $0.5''$ and $0.1''$
thin jet. 
In case B the BC emission is still above
the {\sl Spitzer} sensitivity limit for the two shorter wavelength bands.
However, the existing {\it HST} optical limits cannot be used to argue 
against an $e-p$ jet in this case.
At $\lambda=3.6-8.0 \; \mu m$, 
the most likely band for the BC scattered emission to appear,
the {\sl Spitzer}  angular resolution ($\sim 1''-3''$) is considerably
smaller than the  $\sim 8''$ bridge, and we anticipate that  {\sl Spitzer}
will resolve the BC emission along the bridge.

\section{Discussion}

Existing {\sl HST} limits for PKS 0637-752 already 
disfavor  case A  $e-p$ models.
Additional constraints for  pure $e-p$ jets come from the 
large Lorentz factors required. 
Although values of  $\delta_{min}\sim 30$  are  compatible with the 
 superluminal motions observed in   some blazars 
(e.g. Jorstad et al. 2002),  the number of such 
highly relativistic sources should not overproduce the  
 parent  (misaligned)  population (e.g. Lister 2003).
Additionally,
the large Doppler factors required for   pure $e-p$ jets, suggest that
the actual jets are  over $1$ Mpc long, a value reached by only  the largest
radio galaxies.

An assumption made in our calculations is that the Doppler factor of the 
jet flow in the bridge between the core and the knot is the same as the
Doppler factor of the knot. This can happen if the  flow
does not decelerate substantially at the knot.
This seems to be the case in  PKS 0637-752, where
VLBI observations of superluminal velocities with $v_{app}=17.8\pm 1 \,c $
 in the core of the source (Lovell et al. 2000)
set limits of $\Gamma>17.8$, $\theta<6^{\circ}.4$, in agreement with the
 Doppler factor $\delta=17.4$ derived from minimizing the jet power in an
$e^{\pm}$ jet.

As was first discussed by Schwartz (2002) the X-ray emission due to EC off 
the CMB will remain visible at the same flux level independently of redshift.
This is also the case for the BC  emission.
This suggests an exciting possibility for  jets that have
a very low radiative efficiency past the core: their IR-optical  BC emission 
will be detectable
independent of redshift, and it will be the only observable signature of 
these otherwise invisible jets.

\bibliographystyle{aa}

\begin{thebibliography}{}


\bibitem[{Celotti,  Ghisellini \&  Chiaberge (2001)}]{celotti01}Celotti, A., Ghisellini, G. \&  Chiaberge M. 2001, \mnras,  321, L1


\bibitem[{Chartas et al. (2001)}]{chartas01} Chartas, G. et al. 2001, ApJ, 542, 655

\bibitem[{Dermer & Atoyan (2004)}]{dermer04} Dermer, C D., Atoyan, A.,  ApJ, 
611, L9 

\bibitem[{Georganopoulos et al. (2004)}]{georganopoulos04} Georganopoulos, M.,
Kazanas, D., Perlman, E. S., Stecker, F. W. 2004, ApJ, submitted

\bibitem[{Jorstad et al. (2002)}]{jorstad02} Jorstad, S. G., et al. 2002, ApJ, 556, 738

\bibitem[{Lister (2003)}]{lister03} Lister, M. L. 2003, ApJ, 599, L105

\bibitem[Lovell et al. (2000)]{lovell00} Lovell, J. E. J., et al. 2000, in Astrophysical Phenomena Revealed by Space VLBI, ed. H. Hirabayashi, P. G. Edwards, \& D. W. Murphy (Sagamihara: ISAS), 215


\bibitem[{Reynolds et al. (1996)}]{reynolds91} Reynolds, C. S., Fabian, A. C., Celotti, A. \& Rees, M. J. 1996, MNRAS, 283, 873

\bibitem[{Schwartz et al. (2000)}]{schwartz2000} Schwartz, D. E. et al. 2000, ApJ, 540, L69

\bibitem[Schwartz  (2002)]{schwartz02} Schwartz, D. E.  2002, \apj, 569, L23

\bibitem[{Sikora \& Madejski (2000)}]{sikora00} Sikora, M. \&  Madejski, G. 2000, ApJ, 534, 109

\bibitem[{Tavecchio et al. (2000)}]{tavecchio00}   Tavecchio, F., Maraschi, L., Sambruna, R. \& Urry, C. M., 2000, ApJ, 544, L23
	

\bibitem[{Wardle et al. (1998)}] {wardle98}	 Wardle, J. F. C., Homan, D. C., Ojha, R., Roberts, D. H.  1998, Nature, 395, 457

\end{thebibliography}

\end{document}